\documentclass[useAMS]{mn2e}
\usepackage{psfig}
\usepackage{url}
\usepackage{color}
\usepackage{subfloat}
\usepackage{graphicx}
\usepackage{caption}
\usepackage{amsmath}
\usepackage{breqn}

\def\prd{{\rm Phys. Rev. D}}

\def\aj{{\rm AJ}}
\def\apjs{{\rm ApJS}}

\def\mnras{{\rm MNRAS}}

\def\etal{{\rm et~al.\ }}
\def\hmpc{\;h^{-1}{\rm Mpc}}
\def\hgpc{\;h^{-1}{\rm Gpc}}
\def\invhmpc{\;h\;{\rm Mpc}^{-1}}

\def\hkpc{h^{-1}{\rm kpc}}
\def\kms{{\rm \;km\;s^{-1}}}

\def\kmsmpc{\kms\;{\rm Mpc}^{-1}}
\def\msun{{h^{-1} M_{\odot}}}

\def\simlt{\lower.5ex\hbox{$\; \buildrel < \over \sim \;$}}
\def\simgt{\lower.5ex\hbox{$\; \buildrel > \over \sim \;$}}
\def\rhobar{\bar{\rho}}
\def\bi{\bibitem[]{}}

\title[Gravitational redshifts]{
Gravitational redshifts from large-scale structure
}

\author[R.A.C. Croft]{\parbox{18cm}{
Rupert A.C. Croft$^{1,2}$\thanks{E-mail: rcroft@cmu.edu}
}\vspace{0.3cm}\\
$^{1}$ McWilliams Center for Cosmology, Dept. of Physics, 
Carnegie   Mellon  University, Pittsburgh, PA 15213, USA\\
$^{2}$ Astrophysics, University of Oxford,
Keble Road, Oxford OX1 3RH, UK\\
}
\begin{document}

\pagerange{\pageref{firstpage}--\pageref{lastpage}} \pubyear{2005}

\maketitle

\label{firstpage}

\begin{abstract}
The recent measurement of the gravitational redshifts of galaxies in  
galaxy clusters by Wojtak et al.
has opened a new observational window on dark matter and modified gravity.
By stacking clusters this determination effectively used the line of sight
distortion of the cross-correlation
function of massive galaxies and lower mass galaxies to
estimate the gravitational redshift profile of clusters out
to $4 \hmpc$
Here we use a halo model of clustering
to predict the distortion due to gravitational 
redshifts of the cross-correlation function on 
scales from  $1 \sim 100 \hmpc$. 
We compare our predictions to simulations and use the simulations
to make mock catalogues relevant to current and
future galaxy redshift surveys. Without
formulating an optimal estimator, we
find that the full BOSS survey should be able to
detect gravitational redshifts from large-scale structure at the
$\sim 4\sigma$ level. 
Upcoming redshift surveys will greatly increase the
number of galaxies useable in such studies and 
the BigBOSS and Euclid experiments should be capable of
measurements with precision at the few percent level.
As has been recently pointed out by McDonald, Kaiser and Zhao et al,
other interesting effects
including  relativistic beaming and transverse Doppler shift  can
add additional asymmetric 
distortions to the correlation function. While
these contributions are subdominant to the gravitational redshift
on large scales, they represent additional opportunities to probe gravitational
physics and indicate that many qualitatively new measurements
should soon be possible using large redshift surveys.
\end{abstract}

\begin{keywords}
Cosmology: observations 
\end{keywords}

\section{Introduction}
\label{intro}
In the weak field limit, the gravitational redshift, $z_{g}$ of photons
with wavelength $\lambda$ emitted in a gravitational potential $\phi$
and observed at infinity is given by $z_{g}=\frac{\Delta\lambda}{\lambda}
\simeq \frac{\Delta \phi}{c^{2}}$.
Measurement of $z_{g}$ is one of the fundamental
tests of General Relativity (GR). First measured more that 50 years ago 
for the Earth's gravity in a laboratory setting (Pound \& Rebka 1959),
subsequent determinations have been made in the solar system
(Lopresto \etal 1991) and from spectral line shifts in
red giant stars (e.g., Greenstein \etal 1971). In this paper we will
examine how well the gravitational redshifts caused by the 
largest potential fluctuations in the Universe can be measured using
galaxy redshift surveys.

 Predictions for 
the gravitational redshifts of galaxies in
clusters were computed using analytic models by Cappi (1995, see
also Nottale 1990).
Cappi found that in the most massive ($ \simgt 10^{15} \msun$) clusters the
central galaxy is expected to have a redshift of a few tens of $\kms$
with respect to other cluster members. Kim \& Croft (2004, hereafter
KC04) showed that
instead of using single extremely massive clusters, large galaxy surveys could
be used to make a statistical measurement of the gravitational redshift 
profile.
McDonald (2009, hereafter M09) examined the issue in Fourier
space and pertubation theory, studying the effect of
gravitational redshifts on the large-scale
cross-power spectrum  of different populations
of galaxies.

The first observational determination of galaxy gravitational 
redshifts due to their large-scale environment was made by
Wojtak et al. (2011, hereafter 
W11) using galaxy redshift data from the Sloan Digital
Sky Survey (SDSS). W11 used 125000 galaxies in 7800 galaxy clusters to make
statistical measurement of $z_{g}$ vs. brightest cluster galaxy
distance out to a radius of $4 \hmpc$ ($h=H_{0}/100 \kmsmpc$). 
Their $2.6\sigma$ measurement 
was compared to modelling of the mass distribution from galaxy velocity
dispersions and used to put constraints on modified gravity models.
Dominguez-Romero (2012) also used SDSS data to carry out such a 
test and also found good agreement with GR.

The redshift of a galaxy is a sum of 3 components, the Hubble redshift,
the Doppler redshift from the line of sight 
peculiar velocity and the gravitational 
redshift:
\begin{equation}
cz=H(z)r+v_{\rm pec}+cz_{g},
\label{cz}
\end{equation}
where $H(z)$ is the Hubble parameter
and $r$ the radial distance between
the observer and galaxy in Mpc.
The autocorrelation function of galaxies is often measured as a function
of separation between pairs of galaxies along the line of sight 
($r_{\parallel}$) and across it ($r_{\perp}$). The
$v_{\rm pec}$ and $cz_{g}$ terms in Equation \ref{cz} distort the
correlation function of galaxies in the $r_{\parallel}$ direction. For
the autocorrelation this distortion is symmetric about  
$\Delta(r_{\parallel})=0$. However, more highly clustered galaxies 
(in practice also more massive) lie in deeper potential 
wells  and so have a higher $cz_{g}$ on 
average than less clustered (and less massive)
 galaxies. This means that the cross-correlation
function of two subsets of galaxies with different clustering strengths
will be asymmetric about the $r_{\perp}$ axis. If
the cross-correlation function is centered on the most massive galaxy the
surrounding galaxies will have an relative blueshift. This blueshift 
can be measured as a function of galaxy pair separation and therefore
provide a measure of the scale dependence of the large scale 
gravitational potential around galaxies.

The cross-correlation can be carried out between any subsets of galaxies.
In KC04 and W11 the higher mass subset were brightest cluster galaxies
(BCGs)
and the lower mass subset were other cluster members and nearby 
field galaxies. M09 considered two general populations of galaxies with 
different bias parameters and showed that in Fourier space the
gravitational redshift leads to an imaginary term in the cross-spectrum.
The analysis of M09 was carried out in linear theory, but also included
other terms in the distortion of clustering (related
to the  transformation from observable to proper coordinates, see
Yoo \etal 2009,2011). Our current study will include non-linear scales
and make use of N-body simulations to
formulate and test the simplest measurement scheme. Our aim is to 
be able to unify the two regimes that were considered 
previously, virialized objects and large scale linear structure and
formulate a model and analysis method applicable to all scales.

Because the predicted gravitational redshift of galaxies is typically much 
smaller than the other terms in Equation \ref{cz},
there are several other effects
which need to be considered when devising a scheme to measure
$z_{g}$ from galaxy clustering. Most of these
effects have been considered in the
context of galaxy clusters and are most important in 
virialized regions. For example, Zhao et al. (2012) have shown that
that because the galaxies around the cluster center are not at rest, 
special relativistic time dilation leads to an additional wavelength
shift between those and the center (BCG in the case of a cluster).
Zhao et al. term this the Transverse Doppler effect and point out 
that it will be opposite in sign to $z_{g}$.
Kaiser (2013, hereafter K13) recently explored several more effects, including 
special relativistic beaming which will cause infalling galaxies behind
a cluster to be brighter. This will boost the number above a magnitude limit
behind a cluster compared to in front and give an asymmetry of similar order
to the $z_{g}$ effect. K13 (see also KC04)
also considered the cluster in  the
context of an infalling density field that joins the Hubble flow at large
radii. This and the effects explored in M09 can in principle be modelled 
and marginalized over when looking the effect of gravitational redshifts,
so that we leave a detailed study of their properties to future work. We
give some comments on their likely
relative magnitude and other potential effects in
Section \ref{disc} but in the present paper we concentrate on predictions for
$z_{g}$.

Our our plan for the paper is as follows. In Section
2 we use halo model-inspired fits to the galaxy-mass cross-correlation
to predict the mean relative 
$z_{g}$ between pairs of galaxies in two distinct
populations. We also compute how this $z_{g}$ will act
together peculiar velocities to distort the cross-correlation
function of galaxies from the two populations, focussing
on the residual asymmetry about the $r_{\perp}$
axis. We test our
model with simulations in Section 3. In Section 4 we make simple mock
catalogues relevant to current and future redshift surveys
to estimate error bars which could be expected with
observational data. In Section 5 we summarise our results and
comment on other sources of asymmetric distortion in galaxy clustering.

\section{Large-scale Clustering and gravitational redshifts}
\label{lss}
A way to measure
statistical information
on the large-scale gravitational potential is to examine how
a galaxy correlation function is distorted by gravitational 
redshifts. The theoretical prediction is effectively
the same calculation as carried out for the cluster
redshift profile by Nottale (1990) and later authors,
but on larger scales. The auto-correlation function has a pairwise
symmetry even with $z_{g}$ included, but this is broken for the
cross-correlation function of two populations
of galaxies with different masses (or bias parameters). We then need
to calculate, given the 
higher mass (or bias) galaxy at the origin, what the 
relative blueshift is for the lower mass galaxies
clustered around it.

We take a set of galaxies and split them into two subsets by
mass, labelling the higher mass subset $g1$ and the lower mass
subset $g2$.
After averaging over pairs of $g1$ and $g2$ galaxies, the 
mean gravitational redshift difference between $g2$ galaxies and
$g1$ galaxies separated by  distance $r$  is
\begin{equation}
z_{g}(r)=\frac{G}{c}\int^{r}_{\infty}M_{12}(x)x^{-2} dx
\label{genrel}
\end{equation}
where $M_{12}(r)$ is the mass given by
\begin{equation}
M_{12}(r)=4\pi\rhobar\int^{r}_{0}(\xi_{g1\rho}(x)-\xi_{g2\rho}(x))x^{2} dx.
\end{equation}
Here $\xi_{g1\rho}$ is the $g1$ galaxy-mass cross-correlation function,
$\xi_{g2\rho}$ is the $g2$ galaxy-mass cross-correlation function,
 $\rhobar$ is the mean density of the Universe, $c$ is the
speed of light and $G$ is Newton's gravitational constant.

Equation \ref{genrel} is the prediction for the gravitational
redshift in General Relativity. In order to give a simple example
of what some alternative theories predict for this quantity, we will 
also make use of predictions for the $f(R)$ gravity model used by W11,
derived from one of the models simulated by
Schmidt (2010). For this, model it
was found by Schmidt (2010) that $z_{g}(r)$ in virialized regions should
be multiplied by a factor of 1.33.
We note that $f(R)$ gravity theories can give various values for
the difference between the dynamical and lensing potential (Schmidt 2010)
and that our use of this result is primarily illustrative of
the rough size of possible effects from non-GR models. Also, while we will plot
this $f(R)$ curve on larger than intracluster scales it has not been shown
to be relevant there and in this case acts to guide the eye, illustrating
the effect of a $33\%$ difference from GR.

\begin{figure*} 

\begin{center}
\begin{minipage}[t]{2.5in}
\begin{center}
\includegraphics[width=2.5in,angle=-90]{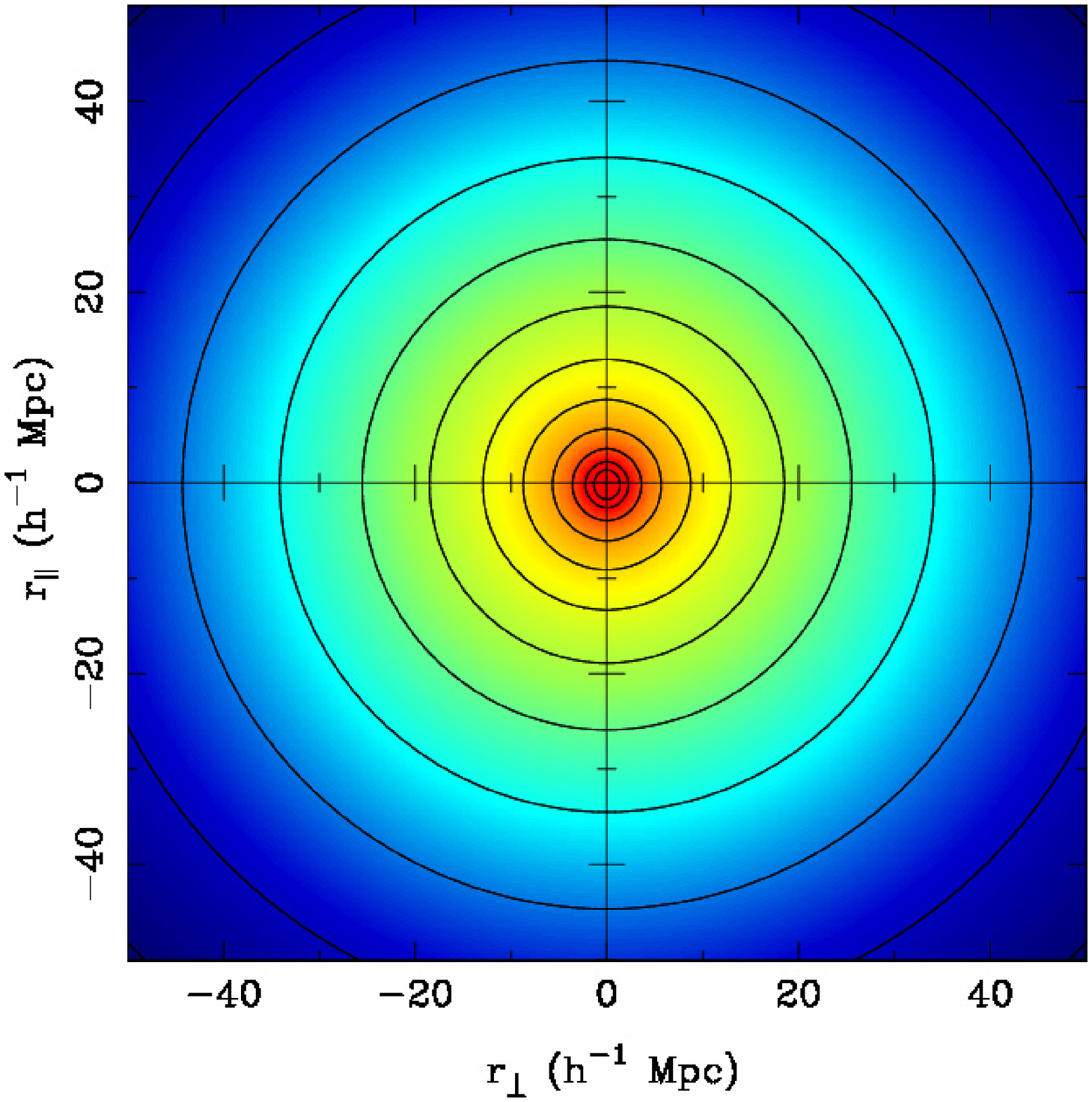}
\vspace{0.3cm}
{(a) Real space}
\end{center}
\end{minipage}
\qquad
\begin{minipage}[t]{2.5in}
\begin{center}
\includegraphics[width=2.5in,angle=-90]{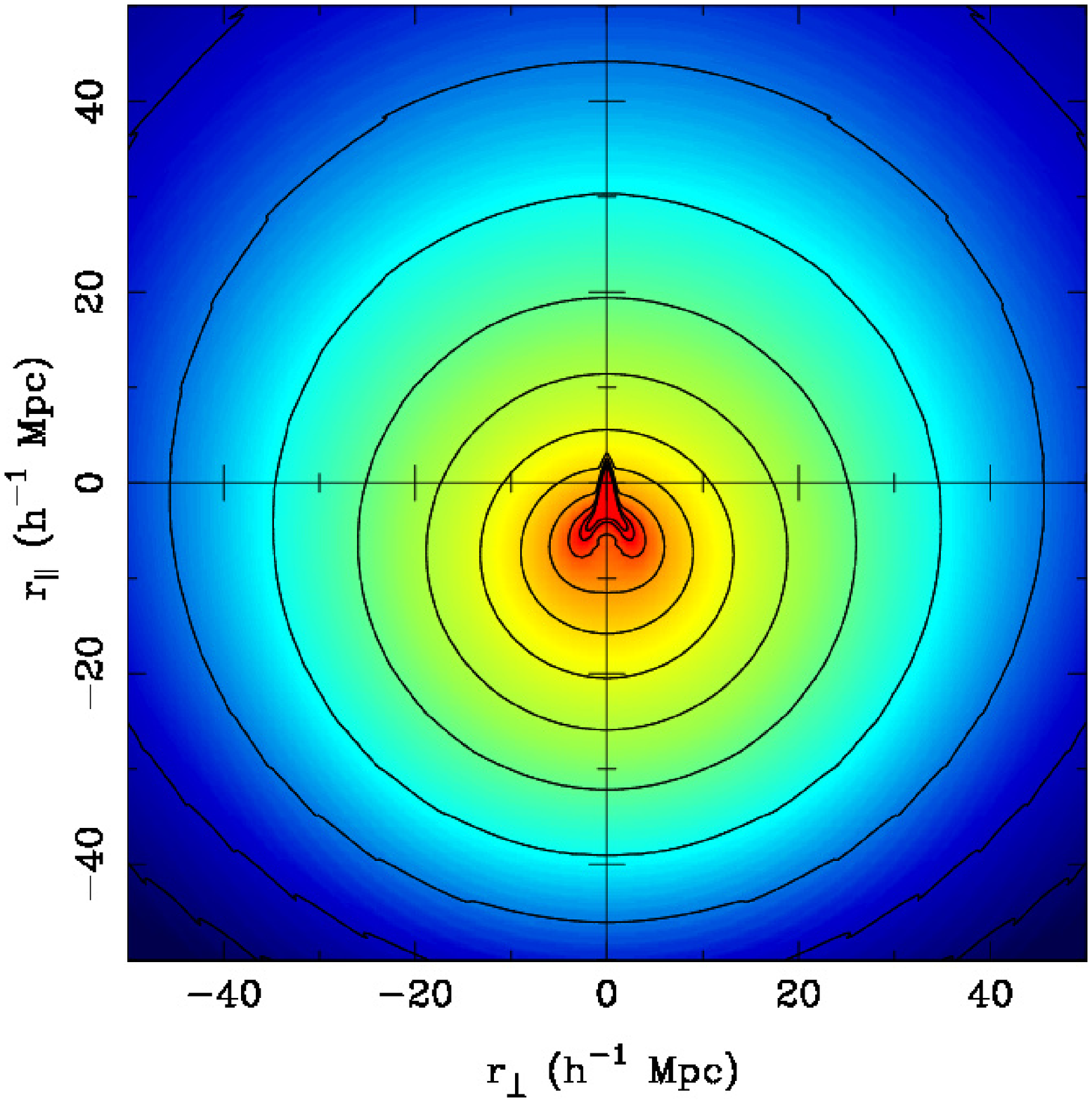}
\vspace{0.3cm}
{(b) With grav. redshifts $\times 500$}
\end{center}
\end{minipage}
\qquad
\end{center}
\begin{center}
\begin{minipage}[t]{2.5in}
\begin{center}
\includegraphics[width=2.5in,angle=-90]{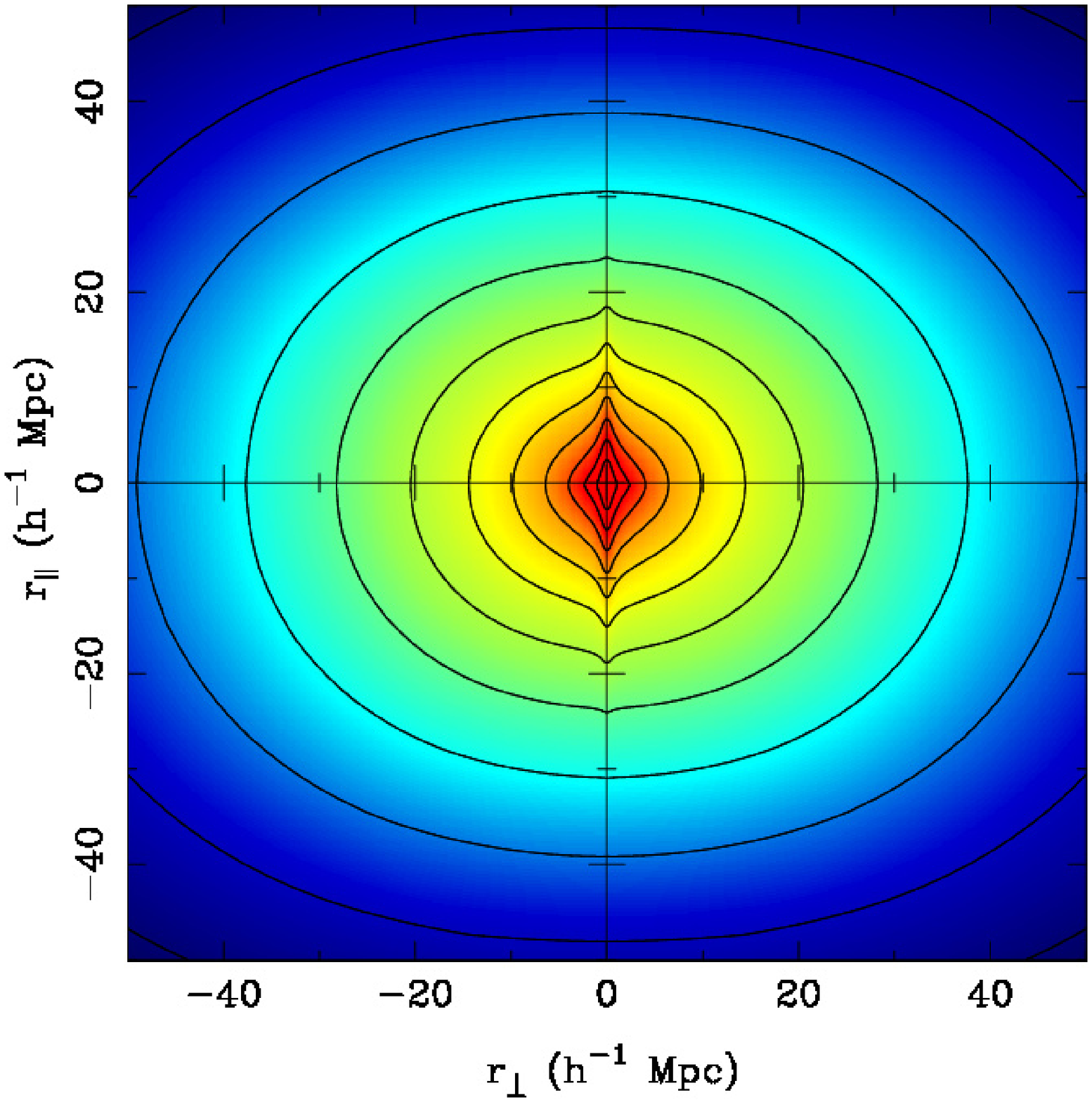}
\vspace{0.3cm}
{(c) With pec. vels.}
\end{center}
\end{minipage}
\qquad
\begin{minipage}[t]{2.5in}
\begin{center}
\includegraphics[width=2.5in,angle=-90]{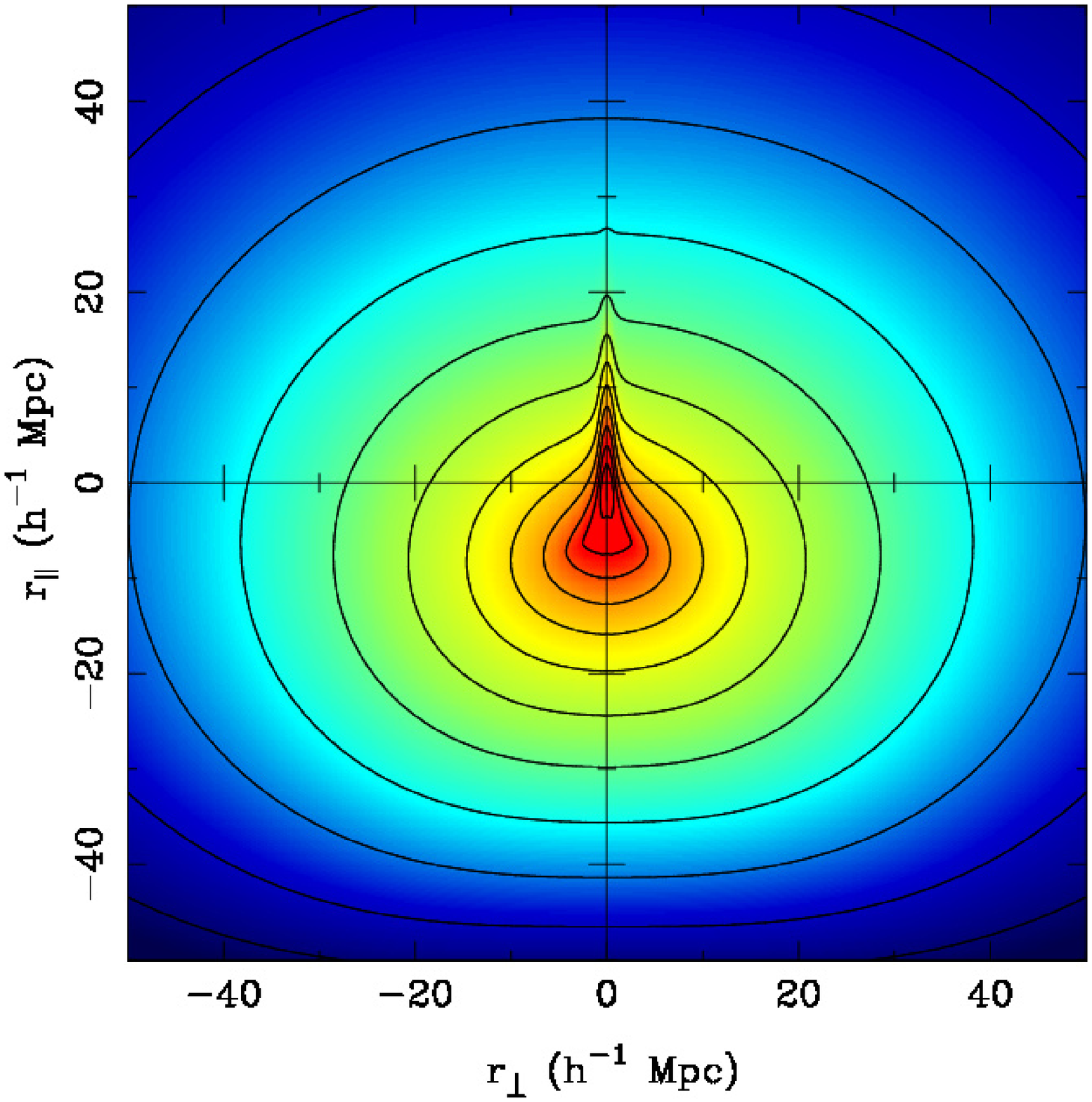}
\vspace{0.3cm}
{(d) With grav. redshifts $\times500$ and pec. vels.}
\end{center}
\end{minipage}
\end{center}
\caption{{Redshift distortions of the cross-correlation function of
two populations of galaxies (high and low mass), using 
a halo model of galaxy clustering (Section \ref{lss}).
We show the effect of peculiar velocities and gravitational redshifts
separately (top right and bottom left panel respectively)} and
together (bottom right). The magnitude of the gravitational
redshift component has been magnified by a factor of $500$ 
in order to make its effect easily visible.}
\label{xi}
\end{figure*}

For $\xi_{g1\rho}$ and $\xi_{g2\rho}$ we use a fit inspired
by the fact that clustering of galaxies and dark matter
can be described by terms which are a combination of internal 
halo structure and large-scale, linear clustering (the halo model 
see e.g., Sheth
and Cooray 2002 for a review). We use
\begin{equation}
\xi_{g\rho}(r) = \rho_{0}\left[\frac{r}{R_{s}}(1+\frac{r}{R_{s}})^{2}\right]
^{-1}+b\xi_{\rm CDM}(r),
\label{eqnfw}
\end{equation}
where we assume that the internal galaxy halo structure follows
the Navarro et al. (1997) relation ($\rho_{0}$ and $R_{s}$ are
free parameters) and 
$\xi_{\rm CDM}$ is the linear theory Cold Dark Matter correlation function
(computed using
Lewis \& Challinor, 2011). Here $b$ is a linear bias parameter relating
galaxy and matter clustering on large scales
and is a monotonically increasing
function of galaxy mass (see e.g. Mo and White 1996).

Having computed $z_{g}(r)$ we use it to distort the
cross-correlation function  of $g1$ and $g2$ galaxies in redshift space,
$\xi_{g1g2}(s_{\perp},s_{\parallel})$.
Here $s_{\perp}$ and $s_{\parallel}$ are respectively the 
distance between pairs of $g1$ and $g2$ galaxies perpendicular and
parallel to the line of  sight, and $s_{\parallel}$ can be either
positive or negative.

 There is also the additional distortion 
of $\xi_{g1g2}(s_{\perp},s_{\parallel})$ due to peculiar velocities.
As in Croft \etal (1999), we model this using a spherical infall model
for large scale flows (taken from Yahil 1985) and a small scale
random velocity dispersion (e.g., Davis and Peebles 1983). 
The infall model is
\begin{equation}
v_{\rm infall}(r)= -\frac{1}{3}\Omega_{0}^{0.6}H_{0}r\frac{\delta(r)}{[1+
\delta(r)]^{0.25}}
\end{equation}
where $\delta(r)$ is the matter overdensity averaged within radius
$r$ of $g1$ galaxies:
\begin{equation}
\delta(r)=\frac{3}{r^{3}}\int^{r}_{0}\xi_{g1\rho}(x)x^{2}dx
\end{equation}
Croft \etal (1999) found that this velocity field model gives a better match
than linear theory for the infall pattern around galaxy clusters. Because
here we are interested in the redshift distortions around massive galaxy halos,
we expect the model to work well in the current context also. This is
borne out in our tests with numerical simulations (Section \ref{sims}).
Because $v_{\rm infall}(r)$ is not expected to describe the virialized
regions of clusters, we follow Croft \etal (1999) and truncate
$v_{\rm infall}(r)$  on small scales by multiplying by an
exponential, $\exp{-(\delta/50)}$.

The random velocity dispersion we use is an exponential model,
so that the distribution function of velocities is
\begin{equation}
f(v)=\frac{1}{\sigma_{12}\sqrt{2}}\exp\left(-\frac{\sqrt{2}|v|}{\sigma_{12}}
\right),
\label{fv}
\end{equation}
where $\sigma_{12}$ is the pairwise velocity dispersion of $g1-g2$
pairs of galaxies, which we assume to be independent of 
pair separation. Based on 
simulation results we take this value to $\sigma_{12}=400 \kms$.

After applying the infall
model, gravitational redshift and 
convolving with $f(v)$, the redshift space 
cross-correlation function is therefore:

\begin{dmath}
\xi_{g1g2}(r_{\perp},r_{\parallel})=\int^{\infty}_{\infty} dv f(v)
\xi_{g1g2,iso}\left(r_{\perp},r_{\parallel}-H_{0}^{-1}(
cz_{g}(r)-r_{\parallel}r^{-1}v_{\rm infall}(r)
-v)\right),
\label{distort}
\end{dmath}
(valid at redshift $z=0$) 
where $\xi_{g1g2,iso}$ is the isotropic (real space) cross-correlation
function of $g1$ and $g2$ galaxies, which we also model using Equation
\ref{eqnfw}.

To make theoretical predictions for $\xi_{g1g2}$ we fit the free parameters
in Equations \ref{eqnfw} and \ref{fv}  ($b$, $\rho_{0}$, $\sigma_{12}$)
using simulations (see Section \ref{sims}).
In Figure \ref{xi} we show contours of $\xi_{g1g2}(r_{\perp},r_{\parallel})$,
illustrating the effects of the different redshift distortions.
We use parameters in Equation \ref{eqnfw} which are appropriate for
halos with mass $>3\times10^{13} \msun$, where the two populations
of galaxies $g1$ and $g2$ are the high and mass low halves of the
set of halos.

Because
the gravitational redshift is so small, for illustrative
purposes we have multiplied $z_{g}$ by
a factor of 500 when making Figure \ref{xi}. This should be borne in mind
when assessing the relative effects shown. The top left panel of
Figure \ref{xi} shows the undistorted, isotropic correlation function 
(Equation \ref{eqnfw}). In panel (b) we can see that the effect of 
gravitational
redshifts without peculiar velocities is to shift the contours
of $\xi_{g1g2}$ downwards, corresponding to a relative blue shift
for the $g2$ galaxies clustered around the $g1$ galaxies. In panel (c)
the peculiar velocity distortion has been applied on its own, resulting
in a distortion of $\xi_{g1g2}$  which has reflection symmetry about the
$r_{\perp}$ axis. The large scale squashing of the contours can be
seen (the Kaiser (1987) effect) as well as the small scale elongation
of  $\xi_{g1g2}$ due to the random velocities. 

In panel (d) we show
gravitational and peculiar velocity redshift distortions together. We can see
that as the effect of Kaiser (1987) infall is to boost the correlation function
overall, this will also enhance the strength of
the asymmetric signal due to gravitational
redshifts. This illustrates that both peculiar velocity
and gravitational redshift distortions will need to be modelled together
in order to make precise constraints on cosmological theories using 
$z_{g}(r)$.

Given a set of $g1$-$g2$ galaxy pair separation measurements, one now needs
to formulate an estimator of the asymmetry of clustering which can
probe $z_{g}(r)$. In the galaxy cluster case (KC04, W11), the pair separations
were binned into cylindrical shell bins, which was appropriate because the
clusters were being treated as distinct objects.
In the current large-scale
structure case, we have decided to instead bin the pairs in spherical
shell bins, and our statistic sensitive to
$z_{g}(r)$ is the mean $r_{\parallel}$ 
position of the pair-weighted centroid of each shell:

\begin{equation}
z^{\rm shell}_{g}(r')=
\frac{\int^{r'+\Delta r'}_{r'} H
r_{\parallel} [1+\xi(r_{\perp},r_{\parallel})]r^{2} dr}
{\int^{r'+\Delta r'}_{r'}  [1+\xi(r_{\perp},r_{\parallel})]r^{2} dr}
\label{eqnest}
\end{equation}

This estimator $z_{g}^{\rm shell}(r)$
will tend to zero at large and small
scales. On small scales this is because the  $z_{g}^{\rm shell}(r)$ shift
cannot be larger than the spherical bin radius. On large scales this is
because the clustering tends towards homogeneity and it is not possible to
detect a blueshift or redshift of a homogeneous set of particles. We will 
explore the exact shape of $z^{\rm shell}_{g}(r)$   in Section \ref{sims}
below. We note that other measures of the asymmetric distortion
of the correlation function could be chosen. For example one could bin galaxy
pairs not in spherical bins but in bins matched to the expected shape of
contours of $\xi_{g1g2}(r_{\perp},r_{\parallel})$. 
Or else one could imagine
carrying out a full fit to the observed $\xi_{g1g2}(r_{\perp},r_{\parallel}$
data varying parameters in our theoretical model. In the present paper
we restrict ourselves
to the simple estimator of distortion  in Equation \ref{eqnest}
and leave exploration of possibly more sensitive measures of $z_{g}(r)$ 
to future work.

\section{Simulation tests}
\label{sims}
We now compare the results of the theoretical predictions for
the $z_{g}$ distortion of Section \ref{lss} to results from numerical
simulations. It should be borne in mind that both the
galaxy-mass cross-correlation function  and the $g1$-$g2$ galaxy 
cross-correlation function enter into the predictions for the
observable quantity (Equation \ref{eqnest} ) and so to make predictions
we will make some simplifying assumptions about galaxy formation. Here we
will associate galaxies with dark matter subhalos and assume that 
applying a dark matter mass cut to a subhalo population is equivalent
to applying a luminosity cut to galaxies.

\subsection{Simulations and galaxies}

We have used the {\small P-Gadget}
(see Springel 2001,2005, Khandai \etal 2011) N-body 
code to run 10 realizations of a $\Lambda$CDM universe. The cosmological
parameters used were : amplitude of mass fluctuations, $\sigma_{8} = 0.8$,
spectra index, $n_{s} = 0.96$,  cosmological constant
parameter $\Omega_{\Lambda} = 0.74$, and 
mass density parameter $\Omega_{m} = 0.26$.
The cubic periodic box side length was $1 \hgpc$,
and the number of particles $768^{3}$ per realization, leading to 
a particle mass of $1.5\times 10^{11}\msun$. 
The gravitational force resolution was $20 \hkpc$,
so that subhalo structure is not well resolved (we return to this below).

The bound structure finder {\small SUBFIND} (Springel
2001) was used to find galaxy sized subhalos. Because the current
relevant redshift surveys (e.g., BOSS, Ahn \etal 2012) are focussed on massive
galaxies, we make two different subhalo samples, one where subhalos
are above a mass limit of $10^{13} \msun$ 
(64 particles) and the other above 
$3\times10^{13} \msun$. 

We compute the gravitational redshift $z_{g}$ 
of the central
particle in each subhalo, and assign that to be the $z_{g}$ associated
with the galaxy inside the subhalo.  We have also tried averaging
the gravitational redshifts of all particles inside each subhalo instead.
We find that while this makes a small difference  to the individual
$z_{g}$ values it does lead to problems with the pairwise differences
between satellite and central galaxies. This is because while the central 
part of a central subhalo may be in a deeper potential well than nearby
satellites, averaging particle $z_{g}$ values over the entirety of the
dark matter
subhalo decreases the redshift and can often lead to a blueshift with
respect to nearby small satellites. This would not happen observationally
as the stellar parts of galaxies are more centrally concentrated than the 
dark matter. We therefore account for this by using only the central particle
to assign $z_{g}$ to the galaxy.

\begin{figure}
\centerline{
\psfig{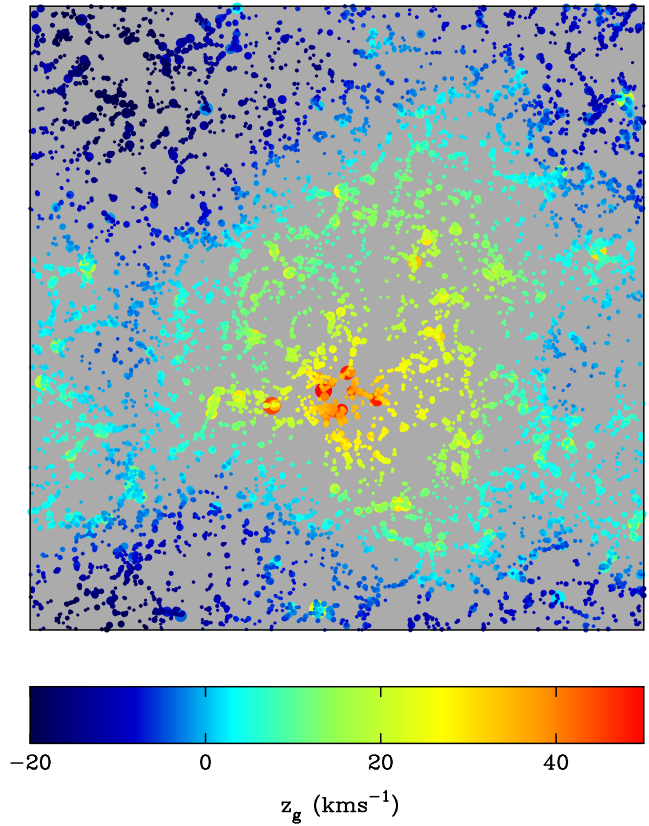}
}
\caption{ Gravitational redshifts of dark matter subhalos in a simulation.     
We show a 10 $\hmpc$ thick slice through a $1 \hgpc$ N-body               
simulation cube at redshift $z=0$.                                             
Dark matter subhalos are shown with symbol size proportional  
to (mass)$^{1/3}$ and color that indicates their gravitational 
redshift (positive numbers for redshift). In this case   
the redshift is not relative to infinity but instead relative to the
mean level in the Universe.                                              
\label{slice}}
\end{figure}

As the gravitational potential is very smooth, the distribution of
gravitational redshifts of galaxies will be also. It is instructive
to plot a slice through the simulation to see the nature of the
$z_{g}$ fluctuations which we will be characterising. We have done this in 
Figure \ref{slice}, where we show $z_{g}$ on a colour scale with galaxy mass
denoted by symbol size. We can see that the most massive galaxies are the
most clustered, as expected, and also that the galaxies in the deepest
potential well (visible near the center of the
slice) are also the most massive. The length scale of the
visible $z_{g}$ fluctuations is extremely large, with the main
potential well covering most of the $\hgpc$ volume. We will be
seeking to measure this structure by measuring $z_{g}$ differences
between pairs of galaxies on scales up to 100 $\hmpc$. On average the
most massive galaxies will tend to be more clustered and have more positive
redshifts (red in Figure \ref{slice}) and this is what we will measure
through the effect of this small shift relative to the nearby lower mass
galaxies.

After making a galaxy catalogue from a simulation
with positions, peculiar velocities
and gravitational redshifts defined as above, we apply a mass cut
to split it into two halves $g1$ and $g2$
 containing an equal number of galaxies.
For the $M > 3\times10^{13}\msun$ sample, for which there
are an average of  $10^{5}$ galaxies in each cubic $\hgpc$,  the mass threshold
which does this is $5.1\times10^{13} \msun$.
In this case, the mean mass of the low mass half ($g1$) is 
$3.8\times10^{13} \msun$ and of the
high mass half ($g2$) $1.2\times10^{14} \msun$.
For the  $M > 3\times10^{13}\msun$ sample a mass
threshold of $1.8\times10^{13}\msun$
split the sample into two halves, of mean mass $1.3\times10^{13}\msun$
and $5.6\times10^{13}\msun$.
We note that defining the $g1$ and $g2$ galaxies to be equal halves
of a sample may not be the optimal split to get a measurement of gravitaional
redshift distortions. For example, in the galaxy cluster (e.g., W11) case the
$g1$ galaxies were all the non-BCG galaxies and so comprised most of the
sample.

\begin{figure}
\centerline{
\psfig{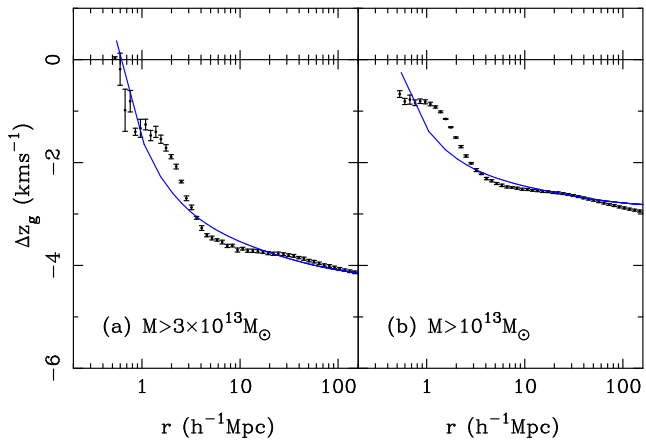}
}
\caption{Mean averaged gravitational redshift difference 
$\Delta z_{g}$ between
pairs of dark matter subhalos as a function of separation. In each panel, the
lower limit for the mass of subhalos in the sample is given. For
each panel, the sample was split into two by mass (high
and low mass halves) and $\Delta z_{g}$ was
computed from mean difference in gravitational
redshifts for galaxies in the two samples as a function of
pair separation. Results for
simulations are shown by points with error bars. The solid curves
show results computed by integrating a simple analytic fit 
to the subhalo-mass cross-correlation function (see Equation \ref{eqnfw})
\label{internal}}
\end{figure}

\subsection{Pairwise $z_{g}(r)$ from simulations}

In the simulations, the gravitational redshift of each galaxy is available,
so that one can directly compute the mean $z_{g}$ difference
between pairs of $g1$ and $g2$ galaxies as a function of pair separation.
While this would not be available observationally, it is a useful first 
test of the prediction for $z_{g}$ from Equation \ref{genrel}. 
For all pairs of  $g1$ and $g2$ galaxies in the simulation we compute
the redshift difference 
$\Delta z_{g}=z_{g,g2}-z_{g,g1}$ and bin the results as a function of 
pair separation. We show the results in Figure \ref{internal}, as
points, with error bars derived from the standard deviation
among the 10 simulation realizations. We also show the predictions
made using Equation \ref{genrel} as a smooth line.
On small scales $r \simlt 0.5 \hmpc$, the
theoretical predictions are relatively sensitive to 
the  internal structure of subhalos, in a regime which 
is not well resolved in the simulations. To deal with this, when 
integrating Equation \ref{eqnfw} we introduce a free parameter
$r_{\rm cut}$ so that $\xi_{g\rho}(r)=\xi_{g\rho}(r_{\rm cut})$
for $r < r_{\rm cut}$, and we adjuste $r_{\rm cut}$ so that the
simulation and theoretical results in Figure \ref{internal} match at 
the small scale end.

As expected, the $g2$ galaxies have a relative blueshift, and this
increases as a function of distance from the $g1$ galaxy, reaching
$\sim  4 \kms$ at $100 \hmpc$
in panel (a) (galaxies above mass $3\times10^{13}\msun$
and $\sim 3 \kms$ for panel (b), the lower mass galaxy catalogue.
There is a knee in the $\Delta z_{g}$ curves at $r \sim 2-3 \hmpc$,
related to the transition in $\xi_{g\rho}$ between the regime
internal to subhalos  and large scale linear cross-correlation 
function. The theoretical predictions broadly follow the 
simulation results, although there are kinks in
the simulation trend at smaller scales. Future work with higher resolution
simulations will be needed to investigate how well the halo model
prediction can be made to fit on these scales. In the present case, we
will see in Section \ref{shellest} that some information on these scales is
washed out by peculiar velocity distortions in any case.

\begin{figure}
\centerline{
\psfig{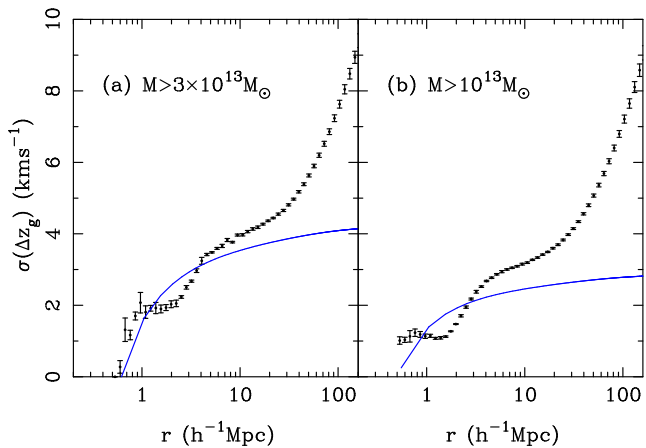}
}
\caption{$\sigma(\Delta z_{g})$, the 
dispersion about mean averaged gravitational redshift difference 
$\Delta  z_{g}$ plotted in 
Figure \ref{internal}.   We again give in  each panel the
lower limit for the mass of halos in the sample. Results for
simulations are shown by points with error bars. The solid curves
not a theoretical prediction for this quantity, but instead are
the theory curves for $z_{g}$  from Figure \ref{internal} to aid
comparison of the two plots.
\label{internaldisp}}
\end{figure}

Equation \ref{genrel} deals with averaged mass and gravitational redshift
profile around galaxies. There will however be 
galaxy to galaxy variation of this quantity, and this will
lead to a dispersion in gravitational redshifts as a function of scale.
Because this dispersion will be symmetric about  $r_{\parallel}=0$, in 
observational data it will be completely hidden by the much larger
symmetric signal due to the dispersion in peculiar velocities and
so unobservable. It is nevertheless instructive to  measure
the dispersion about the mean $\Delta z_{g}$ profile in simulations. One
reason for this is that in Section \ref{mocks} we will make mock catalogues
where model the signal from a much larger number of galaxies by boosting 
$z_{g}$. This will also amplify the contribution of the dispersion in 
$\Delta z_{g}$ so that we need to understand its magnitude.

The results for the dispersion about $\Delta z_{g}(r)$, 
$\sigma (\Delta z_{g}(r))$ are shown in Figure \ref{internaldisp}.
We have also plotted the theory curves for
 $\Delta z_{g}(r)$ from
 Figure \ref{internal} (we have
no theoretical prediction for $\sigma (\Delta z_{g}(r))$ to 
guide the eye. We can see that the dispersion about the mean
profile is approximately equal to the mean profile
until about $\sim 2-5 \hmpc$ and after this it rises relatively steeply
until it is of the order of $\sim 10 \kms$ at $r=100 \hmpc$. It is still
much smaller than the random redshift dispersion from peculiar
velocities, however (which is $\sim 400\kms$).

\subsection{Gravitational redshift estimator}

\label{shellest}

\begin{figure*}
\centerline{
\psfig{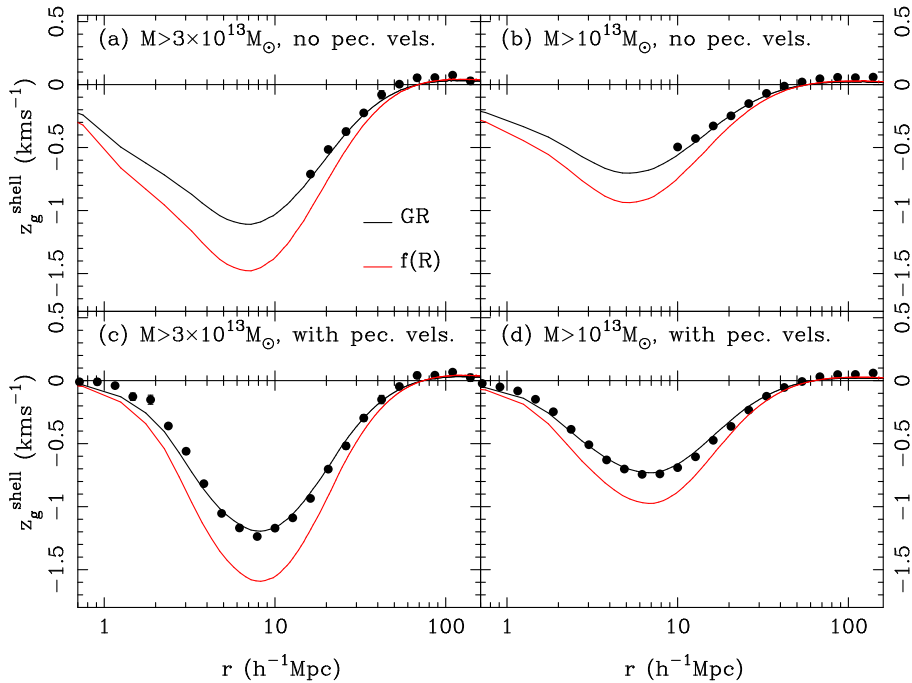}
}
\caption{ 
The gravitational redshift estimator $z^{\rm shell}_{g}(r)$ 
(see Equation \ref{eqnest}) plotted as a function of shell radius $r$.
We show results measured from simulations as points (with error bars
derived from the scatter between 10 realizations). The solid lines show
the halo model inspired predictions of Equations \ref{genrel} and
\ref{eqnest}, for both general relativity (black line) and the f(R) model
of Section \ref{lss}. The top panels show results with no peculiar
velocities added to galaxy redshifts and the bottom panels show the 
results including all terms in Equation \ref{cz} and represent the
observable quantity. The left and right panels show results for 
samples of subhalos with different lower mass limits (given in the 
legend). 
\label{realred}}
\end{figure*}

Having seen that the intrinsic pairwise gravitational redshift 
differences between galaxies in simulations can be modelled, we now
turn to the estimator $z_{g}^{\rm shell}(r)$ of Equation \ref{eqnest}
which can be applied to observational data. Applying the estimator
to simulation data in a periodic volume with a uniform selection function
is equivalent to finding the centroid of shells containing pairs of galaxies
at different separations. In observational data a random catalogue would
need to be used to account for the uniform survey coverage, volume of bins
and so on.

 We would like to show a comparison between
simulation and theory for the $z_{g}^{\rm shell}(r)$ statistic.
One issue is that the $z_{g}$
component of the redshift is extremely small, compared to the other
redshift components, so that averaging over an extremely large number
of simulated galaxies is needed to give a comparison with good signal to
noise. Rather than running hundreds more simulations, we instead choose
to boost the contribution of the $z_{g}$ component by multiplying it
by a large factor (as was done to make Figure \ref{xi}). When computing
the theoretical prediction from Equation \ref{distort} we can also multiply
$z_{g}$ by the same factor (we use a factor of 100 here)
and therefore make a consistent 
comparison. We then divide the final results by this factor in 
order to plot the $z_{g}$ curves with the amplitude they would have
had without boosting them (therefore only the relative
contribution of the noise has been changed).

In Figure \ref{realred}, we show $z_{g}^{\rm shell}(r)$  for the simulations
as points and theoretical predictions as lines, for both galaxy catalogues.
We show the results separately without and with peculiar velocities
(top and bottom panels respectively). When showing the results
without peculiar velocities, the simulations include the dispersion about 
the mean $z_{g}$ profile plotted in Figure \ref{internaldisp}
whereas the theory does not
model this. Because boosting the $z_{g}$ values in the
simulation by a large factor to make
the signal to noise low also boosts the dispersion $\sigma(\Delta(z_{g})$,
this means that comparison with the theory curves is not possible
on small scales. We can see that the theory curves track the large
scale simulation plots that are plotted well. 

The shape of the curves in Figure  \ref{realred} shows the blueshift of
galaxies surrounding more massive ones, which produces a larger and larger
asymmetry in the centroid shift of shells until between 6-9 $\hmpc$ 
when the 
signal decreases. The exact point where this happens will depend
on the shape of the galaxy-mass cross-correlation function. As the
large scale galaxy density field becomes more uniform on large-scales,
this shell centroid blue shift becomes smaller. On largest-scales
($r>50-60 \hmpc$), the shell centroid actually becomes 
slightly redshifted with
respected to the central galaxy.

In the bottom panel of Figure \ref{realred} we show the results in redshift
space, where we can see that because the peculiar velocities are 
dominant in smoothing out the correlation function on small scales that the 
theory and simulations agree reasonably well there also. This is obviously
the situation which will apply when dealing with observations, and so
any remaining differences on small scales between theory and simulations
will be due to a combination of differences in the gravitational 
redshift profile (Figure \ref{internal}) and the peculiar velocity
modelling. 
 
If we compare the top and bottom panels of Figure \ref{realred}, we can
see that the effect of peculiar velocities is to slightly increase
the signal on scales $r > 5 \hmpc$. It can
 be seen that the maximum blueshift of
$g2$ galaxies in shells around $g1$ galaxies is around $1 \kms$. This
is signficantly less than the blueshifts of up to $14 \kms$ seen by W11, and 
shows that by splitting the galaxies into equal halves (of $g1$ and $g2$
galaxies) we make the signal numerically much smaller than 
the redshift profile around rarer galaxies at the centers
of massive clusters. The number of pairs of galaxies
is obviously maximized by our choice, but the signal to noise of
the $z_{g}$ measurement may not be. We leave the question of how best
to partition a galaxy sample to make cross-correlations and also the
best $z_{g}$  estimator to use to future work.

\begin{figure*}
\centerline{
\psfig{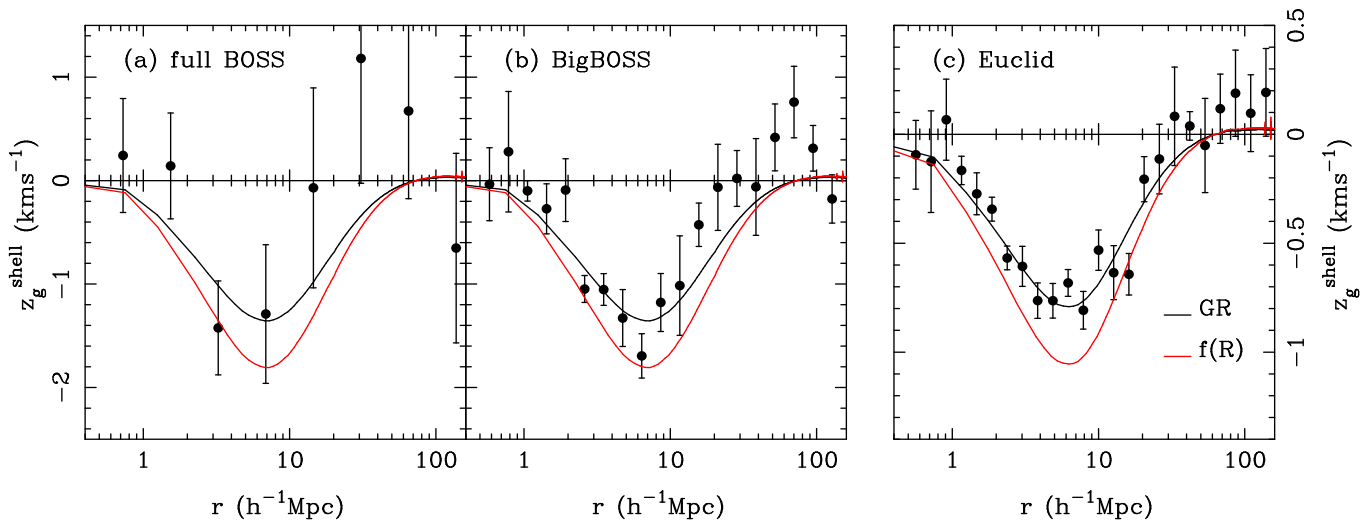}
}
\caption{ 
The points show $z_{g}$                    
 measured  by  observing mock catalogs made from             
simulations. The error bars were computed
from the standard deviation of the measurement
from 10 subsamples in each case.
We show predictions for the                                             
full BOSS survey, BigBOSS and Euclid                  
galaxy redshift surveys.   
The lines are the predictions of Section \ref{lss}
which used a halo model fit to the halo-matter 
cross-correlation function. The black curve shows the
prediction of General Relativity and the red curve
the simple estimate of an $f(R)$ gravity model mentioned in 
Section \ref{lss}.
\label{mocks}}
\end{figure*}

\section{Mock catalogues}
\label{mocks}

In order to make some approximate predictions for  
results that can be expected by measuring $z_{g}^{\rm shell}(r)$
from current and upcoming large redshift surveys, we have made some
very simple mock catalogues from our simulations. The three galaxy redshift
surveys which we deal with are 
\\
(a) the full SDSS/BOSS galaxy survey
(Aihara \etal 2011, Eisenstein \etal 2011) 
which will be complete in 2014 and will contain $1.5 \times 10^{6}$
Luminous Red Galaxies (LRGs) below redshift  $z=0.7$.
\\
(b) the BigBOSS galaxy redshift survey (Schlegel \etal 2009)
which will measure redshifts
for $15 \times10^{6}$ Emission Line Galaxies (ELGs) between redshifts $z=1$ 
and $z=2$
and $3.5 \times10^{6}$ LRGs between redshifts $z=0.2$ and $z=1.0$. The survey
us expected to begin in 2017 and finish in 2022
\\
(c) the Euclid spectroscopic survey (Laureijs \etal 2011)
which will use space-based 
slitless
spectroscopy to measure redshifts 
for between $50-80 \times 10^{6}$ ELGs from  $z=0.5$ to $z=2.0$.
The expected launch date for Euclid is 2020.

As M09 has pointed out, the measurement of $z_{g}$ from the
cross-correlation of two populations of galaxies is limited only
by shot noise and not by cosmic variance. This is the same
property which enables peculiar velocity redshift distortions 
to be measured limited only by shot noise (McDonald \& Seljak 2009).
Therefore the signal to noise 
of a $z_{g}$ measurement in a survey of volume $V$
containing $N_{\rm survey}=\overline{n}V$ galaxies,
scales as $V^{1/2}$, and also as $\overline{n}^{1/2}$
as these quantities are changed (M09). 
As a result when making predictions, to first order
the number of galaxies $N_{\rm survey}$ is the important quantity and 
it is a reasonable approximation to scale
our simulations by a factor to account for the differing shot noise 
resulting from the different numbers of galaxies in each survey.
Although we have $N$-body simulations for
only 10  $(1 \hgpc)^{3}$ volumes, with a maximum $4\times10^5$ galaxies of
mass $>10^{13} \msun$ per volume, this scaling to account for the
number of galaxies allows us to make predictions for surveys
containing many more galaxies.

Other simplifications which we make here are to make the $z-$axis
of the simulation boxes
be the line-of-sight (the plane-parallel approximation),
rather than modelling any actual survey geometrical
effects. We also carry out all measurements at a single redshift $z=0$
(in practice there will also be a dependence on the mean redshift
of the survey e.g., the perturbation growth factor at $z=1$ is 0.62
for the standard cosmology so perturbations in the gravitational
potential will be $38\%$ smaller).
We use the galaxy catalogue with mass limit $m>3\times10^{13}\msun$
to model surveys (a) and (b) and the mass limit  $m>10^{13}\msun$
to model survey (c). For survey (b) we have also not modelled the
ELG and LRG populations separately, but only include in our
mock the $15 \times10^{6}$ ELGs. We also do not include
other asymmetric distortion terms in the correlation function,
such as Transverse Doppler effect (we discuss them in Section \ref{disc}
below).
We leave more sophisticated modelling of observational predictions
to future work. Our aim here is to make a rough assessment
of the detectability of $z_{g}$ signals from large-scale
clustering.

When making mock surveys (a)-(c) we therefore scale
the gravitational redshift $z_{g}$ for galaxies to account
for differing shot noise values. The scaling factor which we
multiply by is $s=\frac{\sqrt{N_{\rm survey}}}
{\sqrt{N_{sim}}}$ where $\sqrt{N_{sim}}$ is the number of galaxies
in all realizations of the simulations added together ($10^{6}$
for the $3\times 10^{13}\msun$ mass limit and $4\times10^{6}$
for the $10^{13}\msun$ mass limit). 
We take $N_{\rm survey}$ to be $1.5\times10^{6}$ for
full BOSS, $15\times10^{6}$ for BigBOSS and $80\times10^{6}$
for Euclid.

In Figure \ref{mocks} we  show the results measured from these
mock catalogues. The error bars show the error on the mean
computed from the standard deviation of 10 subsamples (the 10 simulation
realizations). We can see that the relative blueshift as a function 
of scale should be detectable in the full BOSS survey. The difference
in chi-squared from zero is 16, indicating a 4 $\sigma$ detection.
We have assumed a diagonal covariance matrix however. Future work should
use a larger number of simulations to be able to compute and include
the off diagonal terms.

The
BigBOSS and Euclid surveys should also enable the shape of the curve
as a function of scale to be traced out, with information
on $z_{g}$ up a few tens of $\hmpc$ being accessible. The overall amplitude
of the curves is measurable to $6.5\%$ precision for BigBOSS and $4.0\%$ for
Euclid.

\section{Summary and Discussion}

\subsection{Summary}
Using the halo model, we have made predictions for the relative
gravitational redshifts between pairs of galaxies above and below
a mass threshold as a function of scale. We have shown
how this relates to distortions in the  galaxy cross-correlation
function. Comparing this
to measurements from observational data should enable constraints
to be made on gravity models on large scales. In our theoretical
and simulation study we have found that

(1) The large-scale relative gravitational redshifts between
simulated galaxies can be predicted 
well from the halo model. On small scales, internal to galaxy halos our
comparison is hampered by limited simulation resolution but still gives
good enough predictions for the observable quantity [(2), below].

(2) The gravitational redshifts distort the 
galaxy cross-correlation function so that it is asymmetric about
the $r_{\perp}$ axis. When this asymmetry is quantified by measuring
the pair-weighted centroid shift of spherical shells it peaks
at  a pair separation of $\sim 6-9 \hmpc$.

(3) Peculiar velocities smooth out the correlation function in redshift
space. This has the advantage that after this smoothing
the  small scale clustering which is most difficult to predict does not
have to be known so accurately. It also has the disadvantage that 
peculiar velocity modelling  will form a necessary part of 
accurate interpretation of gravitational redshifts.

(4) Current galaxy redshift surveys such as SDSS/BOSS should be able to detect
large scale gravitational redshifts from the distortion
of the correlation function. Future larger surveys
such as BigBOSS and Euclid should be able to make measurements of
the distortion out to tens of $\hmpc$ scales with a precision good enough
to differentiate between some alternatives to General Relativity.

\subsection{Discussion}
\label{disc}

The measurement of gravitational redshifts of galaxies,
starting with W11 has  effectively opened
a  new window on the distribution of matter in the Universe and
its gravitational effects, complementary
to weak lensing and redshift distortion measurements, but with many 
different dependencies on survey geometry, model parameters and underlying
physics.

Although larger samples of galaxies are required to reach results,
gravitational redshifts have several potential advantages over competing 
probes such as gravitational lensing
or peculiar velocities (see also
M09). For example, lensing effectively gives a
constraint on the projected mass density (the lensing kernel is broad
in redshift), whereas gravitational redshifts are localized in 3D space.
Redshifts also give a constraint on the potential at one instant in
time, rather than an integral over time which is the case for redshift
distortions from peculiar velocities. 

The measurement of $z_{g}$ from large-scale structure and also
the combination of those results with measurements
from galaxy clusters will enable constraints to be put
on cosmological models (for example the amplitude of the signal
in Figure \ref{mocks} will scale with the amplitude of mass
fluctuations, $\sigma_{8}$). By the same
token, the dark matter can be probed, for example by examining the difference
in the predicted profile of galaxy clusters with and without massive neutrinos.
In order to constrain modified gravity scenarios, it will be necessary
to combine the data with other measures of the gravitating matter, for
example through redshift distortions. This was done by W11 on
cluster scales. On large scales, weak lensing and Kaiser (1987) infall
have been used by Reyes \etal 2010 to make such constraints.

We note that the calculation of the
observability of $z_{g}$ effects in large-scale 
clustering by  M09 results in significantly more
pessimistic conclusions than ours (M09 estimates  
$3-5 \sigma$
detectability from a Euclid-scale  experiment). However, W11
detected gravitational redshifts at just under $3\sigma$
significance with only of order $10^{5}$ cluster galaxies.
This indicates that the inclusion of non-linear scale clustering,
(which we have shown can be modelled well in principle) can
dramatically increase the chances of making a precise measurement. As 
an example of this, we have tried recomputing the curves shown 
in Figure \ref{realred} but including only the linear
CDM correlation function term in Equation \ref{eqnfw}. In this case, we find
that the curves on scales $r>20 \hmpc$
in Figure \ref{realred} are similar to the full
Equation \ref{eqnfw} case, as expected (large-scale
clustering is not affected). However on smaller scales the relative blueshift
is reduced by a factor of at least 20 indicating that non-linear clustering
and halo structure are both important to the success of $z_{g}$ measurements.

We have mentioned in Section \ref{intro} that several effects other 
than $z_{g}$ can cause asymmetries in measured galaxy correlation
function. With these effects, one can either decide that they
represent further opportunities to probe physics on large scales,
or one can decide to model and marginalise over them. M09 has
shown for example that when using Equation \ref{cz} the true value of $H(z)$ is
not known, only the observed redshift of the galaxy. The same is
true of the growth parameter determining density fluctuation amplitude.
These and similar issues lead to terms appearing in the prediction
for galaxy clustering which have been calculated in the context of
General Relativity by Yoo \etal (2009,2011). M09 showed that these
terms appear at the same order in wavenumber $k$ as $z_{g}$ terms in the
imaginary part of the cross-spectrum. The calculation of these 
terms relies on 
perturbation theory and their amplitude on the $k\sim 1 \invhmpc$
scales most relevant to the non-linear signal discussed in this paper
is likely to be small, but should be investigated in future work.

Related to this, we remind the reader that we have  only
used the plane parallel
approximation in our simulations and mock catalogues, as well
as neglecting evolution across the simulation volume. As 
K13 has explained, geometrical and other
effects can lead to asymmetries in clustering in redshift
space and these terms need to be dealt with. Around
galaxy clusters, the precise magnitude of geometrical 
effect is related to how galaxies at large distances  
join the Hubble flow. K13 has shown that this
can be large at large impact parameters and will need to
be understood and modelled well. For example, when
dealing with observational data, extremely finely sampled
random catalogs should be made to ensure that the volume-weighted
centroid of shells around galaxies is used to compute 
the statistic plotted in Figure \ref{mocks}. 

Evolution of clustering will lead to asymmetries because the
overall clustering amplitude  on the higher redshift
side of each galaxy will be lower than on the lower redshift side.
This will lead to asymmetries in the correlation function.
These terms are addressed in linear theory by M09 and Yoo \etal (2009)
and in the context of inflow of galaxies into clusters by
K13.

In the virialized regime of clusters, other new terms become
relatively important. For example, Zhao \etal (2012) have shown that the 
Transverse Doppler effect mentioned in Section \ref{intro}
is generically  present
alongside gravitational redshift  because 
the virial theorem relates the
potential to the kinetic energy of galaxies.
In galaxy cluster models inspired by 
the data and measurements of W11, Zhao \etal (2012) find
that the Transverse Doppler
effect (of opposite sign to $z_{g}$) dominates
over $z_{g}$ within the inner $\sim 0.2 \hmpc$ of clusters. At radii
$r > 3 \hmpc$, the contribution drops to $<5\%$ of $z_{g}$ and so it will
be relatively unimportant over most of the scales considered in this paper.

Also important in
clusters is the Doppler beaming effect (K13) which causes galaxies
moving towards the observer to have a higher surface brightness and
hence luminosity. This causes the number of galaxies above a flux
limit to vary and again causes an asymmetry of galaxy density around clusters.
K13 has shown in the case of W11, the flux limit lies at the steep
end of the galaxy luminosity function, so that small changes in 
flux have large effects on the number of galaxies included in the catalogue.
The Doppler beaming effect has the same sign as $z_{g}$ and K13 find that
it is the next most important effect after $z_{g}$ for the W11 parameters. It
also falls off as a fraction of $z_{g}$  at larger radii.

Apart from general and special relativistic, geometry, and clustering
evolution effects on the symmetry of clustering, one can imagine
other uncertainties such as dust extinction (less galaxies seen
behind a galaxy than in front), or lensing magnification (usually
causing the opposite effect, e.g., Hui \etal 2007).
Fang \etal (2011) have examined the
effect of the former on anisotropies in
galaxy clustering and find that its effect
is largest on large scales and along the line of sight.
Incomplete knowlege of galaxy formation can also lead to uncertainty, as 
for example in a realistic observational case we would need to choose
which galaxies to put into high and low mass subsamples using their
luminosities.  Also, when comparing with predictions
from modified gravity models, the difference between large scale mass
distributions
inferred from dynamics (peculiar velocity distortions on)
and from gravitational redshifts will
necessitate some treatment of how galaxies trace mass.

Galaxy redshift surveys are not the only datasets which could
be used to probe the large scale gravitational potential with
gravitational redshifts.
For example, Broadhurst \etal 2000 proposed that with
high resolution X-ray spectroscopy, intracluster gas could be
used  to map out $z_{g}$ 
in large galaxy clusters. More relevant
to the large-scale structure case are 21cm neutral
hydrogen surveys (e.g.,  Peterson \& Suarez 2012). As M09 has pointed out, the
21cm flux field could form
a good contrast to the highly clustered field probed by 
bright galaxies and as $z_{g}$ measurements are differential could
lead to better measurements. In this case, the formalism which we
have outlined in this paper should make it possible to start modelling
the range of scales and densities required.

The current and upcoming generations of large cosmological
surveys will contain enough data that qualitatively new effects
such as gravitational redshift distortions to clustering can be looked for.
With only two published detections so far,  galaxy 
gravitational redshifts are perhaps two decades behind weak lensing
as a field.
Unlike lensing in the early 1990s, however,  surveys two 
orders of magnitude larger than those currently analysed are already
planned for other purposes. These have the promise to bring
gravitational redshifts into the precision cosmology era.

\section*{Acknowledgments}
This work was supported by NSF Award OCI-0749212, the Moore Foundation
and by the Leverhulme Trust's award of a Visiting Professorship at 
the University of Oxford.
The simulations
were performed on facilities
provided by the Moore Foundation in the McWilliams Center for 
Cosmology at Carnegie Mellon University.
I would like to 
thank Volker Springel and Tiziana Di Matteo 
for the use of the {\small P-GADGET} code.
I would also like to thank Nathan Bernier, Andy Bunker, Vincent Desjacques,
Yu Feng, Shirley Ho and Lam Hui for useful discussions
and the hospitality of the Astrophysics Subdepartment in Oxford
where much of this work was carried out.

{}

\end{document}